\documentclass[unnumsec,webpdf,contemporary,large]{oup-authoring-template}

\graphicspath{{Fig/}}

\usepackage{amsmath,amssymb,amsfonts}
\usepackage{hyperref}
\usepackage{xurl}
\usepackage{booktabs}
\usepackage{array}
\usepackage{multirow}
\usepackage{graphicx}

\begin{document}

\journaltitle{Bioinformatics}
\DOI{}
\copyrightyear{}
\pubyear{}
\access{}
\appnotes{Application Note}

\firstpage{1}

\title[AnnotateMissense]{AnnotateMissense: a genome-wide annotation and benchmarking framework for missense pathogenicity prediction}

\author[1,2,$\ast$]{Muhammad Muneeb}
\author[1,2,$\ast$]{David B. Ascher}

\authormark{Muneeb et al.}

\address[1]{\orgdiv{School of Chemistry and Molecular Biology}, \orgname{The University of Queensland}, \orgaddress{\street{Queen Street}, \postcode{4067}, \state{Queensland}, \country{Australia}}}
\address[2]{\orgdiv{Computational Biology and Clinical Informatics}, \orgname{Baker Heart and Diabetes Institute}, \orgaddress{\street{Commercial Road}, \postcode{3004}, \state{Victoria}, \country{Australia}}}

\corresp[$\ast$]{Corresponding author: David B. Ascher, Email: \href{mailto:d.ascher@uq.edu.au}{d.ascher@uq.edu.au}}

\abstract{
\textbf{Motivation:}
Missense variant interpretation remains challenging because pathogenicity depends on heterogeneous evidence, including population frequency, evolutionary conservation, transcript context, amino acid substitution severity, prior pathogenicity predictors and protein-language-model-derived features. Although these resources are individually useful, there remains a need for reproducible workflows that integrate them at genome-wide scale, benchmark their contribution and provide accessible outputs for downstream research.
\textbf{Results:}
We present \textbf{AnnotateMissense}, a scalable annotation, benchmarking and genome-wide prediction framework for missense variant interpretation. AnnotateMissense integrates chromosome-wise hg38 missense variants derived from dbNSFP v5.1 with ANNOVAR-based gene-, region- and filter-based annotations, dbNSFP transcript and protein descriptors, AlphaMissense scores, ESM-derived features, conservation metrics, population-frequency variables, established pathogenicity predictors and engineered amino acid/codon-context features. Using 132{,}714 ClinVar-labelled missense variants, we benchmarked machine-learning and deep-learning models under controlled feature configurations. The full 303-feature benchmark set achieved the strongest performance with XGBoost, reaching mean Matthews correlation coefficient (MCC) = 0.9411 and ROC-AUC = 0.9950 across stratified five-fold cross-validation. Restricted naive and location-oriented feature sets achieved substantially lower best MCC values of 0.4989 and 0.5113, respectively. Circularity-controlled ablations showed that removing prior-predictor, population-frequency and clinically overlapping evidence reduced performance, whereas excluding AlphaMissense and ESM-derived features alone had minimal effect. Temporal ClinVar validation on newly observed pathogenic/benign variants achieved MCC = 0.7613, accuracy = 0.8798 and F1-score = 0.8750. The final genome-wide model was applied to 90{,}643{,}830 hg38 missense variants to generate AnnotateMissense pathogenicity scores and binary prediction labels.
\textbf{Availability and implementation:}
Source code, workflow scripts and command files are available at \url{https://github.com/MuhammadMuneeb007/CAGI7_Annotate_All_Missense}. Genome-wide prediction outputs and the compressed DuckDB database are available at \url{https://doi.org/10.5281/zenodo.19981867}.
\textbf{Supplementary information:}
Supplementary data are available online.
}


\maketitle

\section{Introduction}

Missense variants are among the most clinically important classes of protein-altering genetic variation, but their interpretation remains difficult because functional impact depends on multiple and incompletely overlapping evidence layers. A variant may be rare in population databases, affect an evolutionarily conserved residue, alter amino acid physicochemical properties, occur in an important transcript or protein context, or be supported by prior clinical observations. No single evidence source is sufficient across all genes, proteins and disease settings, creating a persistent challenge for rare disease diagnosis, genome interpretation and large-scale variant prioritisation~\cite{Ng2003sift,Adzhubei2010polyphen2,Richards2015acmg,Landrum2018clinvar,Karczewski2020gnomad}.

Many computational resources have been developed to prioritise missense variants, including conservation-based predictors, ensemble pathogenicity scores, genome-wide annotation frameworks and recent protein-language-model-derived resources~\cite{Kircher2014cadd,Ioannidis2016revel,Feng2017bayesdel,Li2022metarnn,Cheng2023alphamissense,NEURIPS2021_f51338d7}. These methods provide complementary information, but they differ in score directionality, coverage, missingness, output type, training data and genome-build compatibility. This makes large-scale integration and systematic benchmarking difficult, particularly when some predictors may have been trained on ClinVar, HGMD or related clinical variant databases.

Here, we present \textbf{AnnotateMissense}, a reproducible framework for genome-wide missense annotation, feature integration, benchmarking, validation and prediction. Starting from 90{,}643{,}830 hg38 missense single-nucleotide variants derived from dbNSFP v5.1 through the CAGI7 Annotate-All-Missense challenge, AnnotateMissense integrates ANNOVAR- and dbNSFP-derived annotations, AlphaMissense scores, ESM-derived features, population-frequency variables, conservation scores, established pathogenicity predictors and engineered biological features~\cite{Wang2010annovar,Liu2020dbnsfp,Cheng2023alphamissense,NEURIPS2021_f51338d7}. We benchmarked multiple machine-learning and deep-learning models using ClinVar-labelled missense variants, quantified feature-set contributions through controlled ablation analyses, evaluated prospective concordance using temporal ClinVar validation and generated genome-wide pathogenicity scores as a public research resource. AnnotateMissense is intended as a scalable annotation, benchmarking and research-prioritisation framework, not as a standalone clinical classification system.

\section{Materials and methods}

\subsection{Genome-wide annotation and feature integration}

AnnotateMissense was developed using chromosome-wise hg38 missense variant files derived from dbNSFP v5.1. These files comprised 90{,}643{,}830 missense single-nucleotide variants and included genomic, transcript and protein-level descriptors. Chromosome-level variant files were converted into ANNOVAR-compatible input format and annotated using gene-based, region-based and filter-based resources~\cite{Wang2010annovar,Liu2020dbnsfp}. The resulting outputs were merged with dbNSFP-derived transcript and protein context, AlphaMissense scores, ESM-derived missense effect predictions, population-frequency variables, conservation scores, established pathogenicity predictors and engineered amino acid/codon-context features~\cite{Cheng2023alphamissense,NEURIPS2021_f51338d7}.

The integrated feature space included raw annotation variables, transformed numeric variables, categorical encodings, consensus-voting features, amino acid physicochemical change features, codon-composition variables, nucleotide substitution descriptors and selected interaction features. Direct ClinVar-derived fields not intended for prediction were excluded to reduce target leakage. Full annotation commands, database categories, feature construction details and retained/excluded feature summaries are provided in Supplementary Material~1, Supplementary Sections~1--3, and Supplementary Material~2, Sheets~S1--S2.

\subsection{ClinVar benchmark and model evaluation}

ClinVar-labelled missense variants were extracted from the chromosome-level annotation tables~\cite{Landrum2018clinvar}. Pathogenic and likely pathogenic variants were assigned to the positive class, whereas benign and likely benign variants were assigned to the negative class. Variants annotated as uncertain significance, conflicting, ambiguous or unsupported were excluded from the primary supervised benchmark. The resulting benchmark contained 132{,}714 missense variants, comprising 76{,}804 pathogenic/likely pathogenic and 55{,}910 benign/likely benign variants.

Models were evaluated using stratified five-fold cross-validation. Feature preprocessing was fitted within each training fold and applied to the corresponding held-out test fold. XGBoost, Random Forest, FLAML AutoML, TabNet, a PyTorch deep neural network and a TensorFlow deep neural network were benchmarked across controlled feature configurations~\cite{Chen2016xgboost,Breiman2001rf,Arik2021tabnet}. Matthews correlation coefficient (MCC) was used as the primary performance metric, with ROC-AUC, accuracy and related classification metrics used as secondary measures~\cite{Matthews1975mcc,Chicco2020mcc}.

Dataset numbering was fixed throughout the manuscript and supplementary material. Dataset~2 refers to the full 303-feature benchmark matrix, Dataset~3 to the 41-feature naive feature set, Dataset~4 to the 56-feature location-oriented feature set, and Datasets~5--8 to circularity-controlled ablation configurations. Dataset~5 excluded prior pathogenicity predictors and population-frequency metrics. Dataset~6 excluded features derived from tools trained on ClinVar, HGMD or overlapping clinical databases. Dataset~7 excluded AlphaMissense and ESM-derived features. Dataset~8 retained only engineered biological sequence-derived features. Full model settings, feature-configuration definitions and benchmark tables are provided in Supplementary Material~1, Supplementary Sections~3--6.

\subsection{Temporal validation and biological use cases}

To assess generalisation beyond the original ClinVar-derived benchmark, temporal validation was performed using variants present in a newer ClinVar release but absent from the older annotation-derived ClinVar set used during model development. Variants were matched by chromosome, position, reference allele and alternate allele. Only strict pathogenic/benign categories were retained for evaluation, and uncertain or ambiguous categories were excluded. AnnotateMissense predictions were evaluated using the final categorical prediction label. Full temporal validation results are provided in Supplementary Material~1, Supplementary Section~10, and Supplementary Material~2, Sheet~S4.

Biological utility was assessed using two use-case analyses. First, AnnotateMissense scores were evaluated for ClinVar variants of uncertain significance (VUS) stratified by gnomAD missense constraint Z-score~\cite{Karczewski2020gnomad}. Second, pairwise discordance analysis was performed against established pathogenicity predictors to identify variants where AnnotateMissense provided complementary classifications relative to existing methods. Detailed VUS prioritisation and discordance-analysis methods are provided in Supplementary Material~1, Supplementary Sections~11--12.

\section{Results and discussion}

\subsection{AnnotateMissense provides a genome-wide missense annotation and prediction resource}

\begin{figure*}[!ht]
\centering
\includegraphics[width=\textwidth]{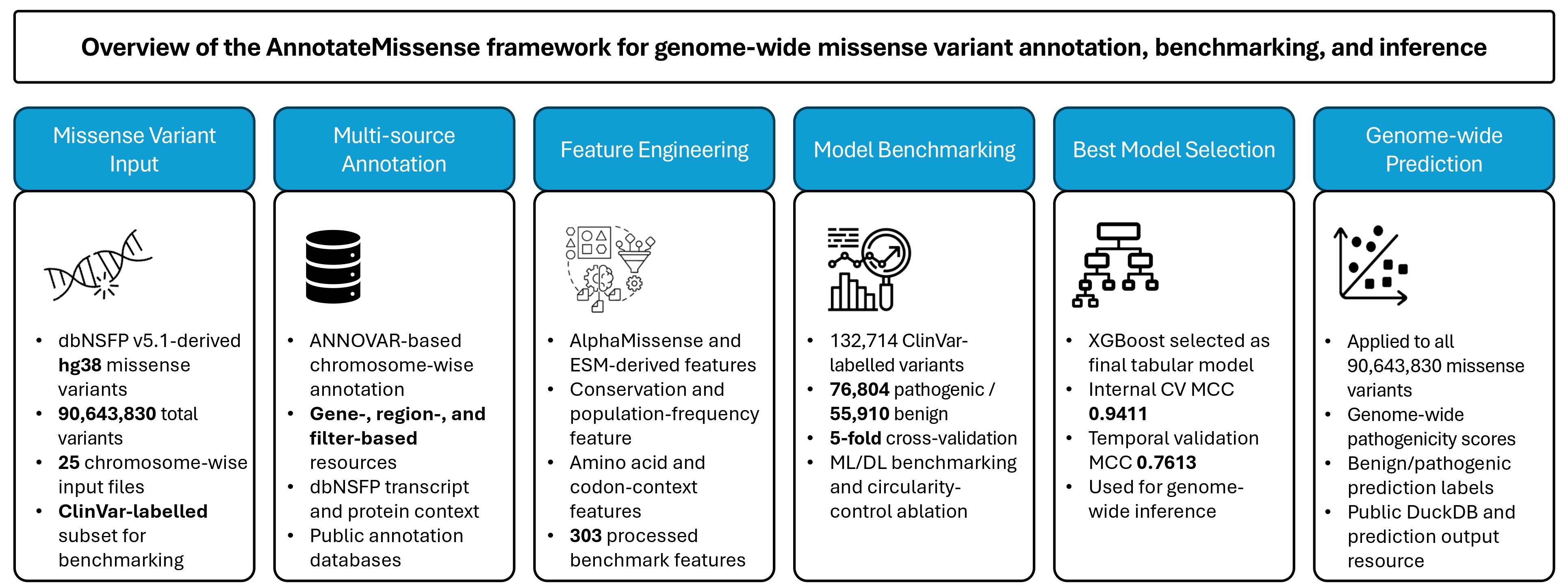}
\caption{
Overview of the AnnotateMissense framework for genome-wide missense variant annotation, benchmarking, validation and inference. Genome-wide hg38 missense variants derived from dbNSFP v5.1 were processed as chromosome-wise input files and annotated using ANNOVAR-based gene-, region- and filter-based resources. The resulting annotation tables were integrated with dbNSFP transcript and protein context, AlphaMissense scores, ESM-derived features, conservation metrics, population-frequency variables, prior pathogenicity predictors and engineered amino acid/codon-context features. ClinVar-labelled missense variants were used for supervised benchmarking across machine-learning and deep-learning models, with circularity-controlled ablation analyses used to assess the contribution of different evidence classes. XGBoost was selected as the final tabular model based on internal cross-validation performance, achieving an internal cross-validation MCC of 0.9411 and temporal ClinVar validation MCC of 0.7613. The final model was applied genome-wide to generate AnnotateMissense pathogenicity scores and binary prediction labels for 90{,}643{,}830 missense variants. The generated DuckDB database and prediction output files are provided as a public research resource.
}
\label{fig:annotatemissense_workflow}
\end{figure*}

AnnotateMissense generated an integrated annotation, benchmarking and prediction resource for 90{,}643{,}830 hg38 missense variants (Fig.~\ref{fig:annotatemissense_workflow}). The workflow links chromosome-wise missense variant inputs to ANNOVAR-based annotations, dbNSFP transcript/protein descriptors, AlphaMissense scores, ESM-derived features, conservation metrics, population-frequency variables, established pathogenicity predictors and engineered biological variables. The final outputs include genome-wide AnnotateMissense pathogenicity scores, binary prediction labels, compressed prediction files and a queryable DuckDB database.

\subsection{Multi-source annotation improves missense classification}

Across all supervised benchmark configurations, the full multi-source feature set produced the strongest performance. On Dataset~2, XGBoost achieved the highest overall performance, with mean test MCC = 0.9411 and mean test ROC-AUC = 0.9950 across stratified five-fold cross-validation (Table~\ref{tab:main_summary}). FLAML AutoML, Random Forest, TabNet and dense neural-network models also performed strongly on the full feature set, indicating that the integrated annotation matrix contained robust predictive signal across model families.

Performance dropped substantially when the feature space was restricted. Dataset~3, the naive feature set, achieved a best MCC of 0.4989, while Dataset~4, the location-oriented feature set, achieved a best MCC of 0.5113. Dataset~8, which retained only engineered biological sequence-derived features, achieved a best MCC of 0.5120. These results indicate that conservation, functional annotation and local sequence-derived features alone were insufficient to recapitulate the performance of the full multi-source feature space.

Because several input features were derived from tools that may have been trained on ClinVar, HGMD or overlapping clinical variant databases, we performed circularity-controlled ablation analyses. Removing prior pathogenicity predictors and population-frequency metrics reduced performance, with Dataset~5 achieving a best MCC of 0.7582. Removing a broader set of clinically overlapping features produced a similar reduction, with Dataset~6 achieving a best MCC of 0.7326. In contrast, removing AlphaMissense and ESM-derived features alone had minimal impact, with Dataset~7 achieving a best MCC of 0.9409. These findings show that AnnotateMissense performance is driven by broad integration of heterogeneous evidence sources, while also highlighting the importance of interpreting benchmark performance in the context of predictor dependence and potential training-data overlap.

\begin{table*}[!t]
\centering
\caption{
Summary of key AnnotateMissense benchmark, ablation, temporal validation and genome-wide inference results. Full model-level results, confidence intervals, comparator analyses and machine-readable outputs are provided in Supplementary Material~1 and Supplementary Material~2.
}
\label{tab:main_summary}
\small
\begin{tabular}{llll}
\toprule
\textbf{Analysis} & \textbf{Dataset / variants} & \textbf{Feature setting} & \textbf{Main result} \\
\midrule
Internal benchmark & 132{,}714 ClinVar variants & Dataset~2, full benchmark & XGBoost MCC = 0.9411; ROC-AUC = 0.9950 \\
Naive feature benchmark & 132{,}714 ClinVar variants & Dataset~3, naive features & Best MCC = 0.4989 \\
Location-oriented benchmark & 132{,}714 ClinVar variants & Dataset~4, location-oriented features & Best MCC = 0.5113 \\
NoPriorPredictors ablation & 132{,}714 ClinVar variants & Dataset~5 & Best MCC = 0.7582 \\
NoClinicalFeatures ablation & 132{,}714 ClinVar variants & Dataset~6 & Best MCC = 0.7326 \\
NoAlphaMissenseESM ablation & 132{,}714 ClinVar variants & Dataset~7 & Best MCC = 0.9409 \\
EngineeredBiologicalOnly ablation & 132{,}714 ClinVar variants & Dataset~8 & Best MCC = 0.5120 \\
Temporal ClinVar validation & 298{,}353 matched new variants & Final categorical prediction & MCC = 0.7613; F1 = 0.8750 \\
Genome-wide inference & 90{,}643{,}830 hg38 missense variants & Final genome-wide model & Public scores, labels and DuckDB resource \\
\bottomrule
\end{tabular}
\end{table*}

\subsection{Temporal validation and biological use cases support prioritisation utility}

Temporal ClinVar validation was performed using newly observed pathogenic/benign variants not present in the older annotation-derived ClinVar set used during model development. AnnotateMissense produced categorical predictions for 298{,}353 newly observed pathogenic/benign variants, comprising 138{,}441 pathogenic and 159{,}912 benign variants. In this available-set temporal validation, AnnotateMissense achieved MCC = 0.7613, accuracy = 0.8798, sensitivity = 0.9070, specificity = 0.8563, precision = 0.8453 and F1-score = 0.8750. These results support prospective concordance with later ClinVar classifications, although they should be interpreted as temporal agreement with clinical variant labels rather than independent clinical validation.

AnnotateMissense was further evaluated in biological use-case analyses. Among 49{,}990 ClinVar missense VUS, variants in missense-intolerant genes showed higher predicted pathogenicity scores than variants in missense-tolerant genes. VUS in genes with missense constraint Z-score greater than 3 had a mean AnnotateMissense score of 0.631, compared with 0.548 for VUS in genes with missense constraint Z-score less than 0. The proportion of high-priority VUS with score greater than 0.9 followed the same gradient, supporting the biological relevance of the prioritisation scores.

Pairwise discordance analyses showed that AnnotateMissense was broadly concordant with established pathogenicity predictors while providing complementary classifications for a subset of variants. On discordant variants, AnnotateMissense showed the strongest complementary resolution against ESM1v, agreeing with ClinVar labels for 67.9\% of variants where the two methods disagreed. These analyses support the use of AnnotateMissense as a research-prioritisation framework for variant triage, benchmarking and hypothesis generation.

\subsection{Framework interpretation and limitations}

AnnotateMissense provides a scalable framework for genome-wide missense annotation, benchmarking, validation and prediction. The main finding is that high-performing missense classification requires integration of heterogeneous evidence sources. Models trained on the full multi-source feature set substantially outperformed restricted feature configurations based on conservation, functional annotation or local sequence-derived features alone. This supports the view that missense pathogenicity is best prioritised using combined evolutionary, biochemical, population, transcript-level and computational-predictor evidence.

The ablation analyses are central to the interpretation of the framework. Performance decreased when prior predictors, population-frequency variables and clinically overlapping evidence were removed, demonstrating that these evidence layers contribute strongly to benchmark performance. Therefore, AnnotateMissense should not be interpreted as an entirely independent clinical truth model. Rather, it is a transparent multi-source annotation and prioritisation framework that integrates existing evidence, quantifies feature-set contributions and provides scalable genome-wide inference outputs for research use.

Predictions generated by AnnotateMissense should be interpreted as prioritisation scores for research, benchmarking and variant triage, not as standalone clinical classifications. Clinical interpretation requires expert review and integration with disease context, inheritance, segregation, functional evidence and current clinical guidelines.

\section{Availability and implementation}

The AnnotateMissense source code, workflow scripts, command files, feature-integration scripts, benchmarking code, temporal validation scripts, biological use-case analyses and genome-wide inference scripts are available at:

\url{https://github.com/MuhammadMuneeb007/CAGI7_Annotate_All_Missense}

The generated genome-wide AnnotateMissense prediction resource is available from Zenodo at:

\url{https://doi.org/10.5281/zenodo.19981867}

The Zenodo record includes the compressed DuckDB database \texttt{variants.duckdb.gz} and the compressed final prediction/output table \texttt{UQ\_BioSig\_model\_Final.tsv.gz}. AnnotateMissense is provided for research prioritisation, benchmarking and variant triage, and should not be interpreted as a standalone clinical classification system.

\section{Supplementary data}

Supplementary Material~1 is available online and contains the extended annotation workflow, representative ANNOVAR commands, dataset definitions, feature configuration details, model training settings, benchmark and ablation results, temporal ClinVar validation, VUS prioritisation, discordance analysis and data/resource information.

Supplementary Material~2 is available online as a machine-readable workbook. Sheet~S1 contains database information, Sheet~S2 contains feature information, Sheet~S3 contains prediction correlation and comparator results, and Sheet~S4 contains temporal ClinVar validation results.

\section{Author contributions statement}

M.M. wrote the first draft of the manuscript and wrote, tested, and documented the code. M.M. analysed the results. D.A. reviewed and edited the manuscript. All authors contributed to the methodology.
\section{Acknowledgments}

D.B.A. is supported by an NHMRC Investigator Grant (GNT2041888).
\section{Conflict of interest}

The authors declare no competing interests.

\section{Data availability}

The primary genome-wide missense variant input used in this study was derived from the CAGI7 Annotate-All-Missense challenge input based on dbNSFP v5.1. The generated AnnotateMissense genome-wide prediction outputs and compressed DuckDB database are available from Zenodo at \url{https://doi.org/10.5281/zenodo.19981867}. Source code and workflow scripts are available at \url{https://github.com/MuhammadMuneeb007/CAGI7_Annotate_All_Missense}.

\bibliographystyle{unsrt}
\bibliography{reference}

\end{document}


\maketitle

\section*{Overview}

This document provides extended supporting material for the AnnotateMissense manuscript. 
The main article reports the primary findings, while this supplementary file provides additional detail on the genome-wide annotation workflow, feature engineering, dataset construction, circularity-controlled feature configurations, model training, temporal ClinVar validation, biological use-case analyses, and extended benchmark tables.

The dataset numbering used throughout this supplementary file is fixed as follows: Dataset~2 corresponds to the full benchmark feature set, Dataset~3 corresponds to the naive feature set, Dataset~4 corresponds to the location-oriented feature set, and Datasets~5--8 correspond to circularity-controlled ablation configurations.

\tableofcontents
\newpage

\section*{Supplementary Section 1. Genome-wide variant annotation workflow}
\addcontentsline{toc}{section}{Supplementary Section 1. Genome-wide variant annotation workflow}

The AnnotateMissense workflow was developed for large-scale missense variant annotation, feature integration, benchmarking, and genome-wide pathogenicity prediction. 
The starting dataset was the CAGI7 \textit{Annotate-All-Missense} challenge input, derived from dbNSFP v5.1 and distributed as chromosome-divided files across the hg38/GRCh38 reference genome. 
These files comprised 90,643,830 missense single nucleotide variants and included genomic, transcript, and protein-level descriptors.

Chromosome-divided missense variant files were converted into ANNOVAR-compatible input files containing chromosome, genomic position, reference allele, and alternate allele information. 
ANNOVAR was then used to annotate variants across gene-based, region-based, and filter-based resources. 
Representative commands are shown below.

\subsection*{Section 1.1. Representative ANNOVAR command using RefSeq gene annotation}

\begin{lstlisting}[language=bash]
perl annovar/table_annovar.pl \
  AnnovarInputFiles/chr1.annovar_input \
  annovar/humandb/ \
  -buildver hg38 \
  -outfile Annovar_Output_Files/chr1_refGene_results \
  -remove \
  -protocol refGene \
  -operation g \
  -nastring . \
  -csvout
\end{lstlisting}

\subsection*{Section 1.2. Representative ANNOVAR command using Ensembl gene annotation}

\begin{lstlisting}[language=bash]
perl annovar/table_annovar.pl \
  AnnovarInputFiles/chr1.annovar_input \
  annovar/humandb/ \
  -buildver hg38 \
  -outfile Annovar_Output_Files/chr1_ensGene_results \
  -remove \
  -protocol ensGene \
  -operation g \
  -nastring . \
  -csvout
\end{lstlisting}

After ANNOVAR execution, protocol-level outputs were merged on a chromosome-wise basis to create unified annotation tables for downstream analysis. 
The merged files were augmented with additional variant and protein context, including amino acid substitution information, transcript-level descriptors from dbNSFP, AlphaMissense pathogenicity scores, and ESM-derived missense effect predictions.

\begin{table}[H]
\centering
\caption{Major annotation sources integrated into the AnnotateMissense workflow.}
\label{tab:s1_annotation_sources}
\small
\begin{tabularx}{\textwidth}{p{3.5cm}X}
\toprule
\textbf{Resource group} & \textbf{Examples of integrated information} \\
\midrule
Gene and transcript annotation & RefSeq, Ensembl, UCSC Known Gene, CCDS, GENCODE, exon function annotations, transcript consequence annotations \\
Population frequency & gnomAD, ExAC, 1000 Genomes, population-specific allele frequencies, allele rarity metrics \\
Conservation & PhyloP, PhastCons, GERP++, SiPhy and related vertebrate/mammalian conservation features \\
Prior pathogenicity predictors & SIFT, SIFT4G, PolyPhen-2, CADD, DANN, REVEL, MetaSVM, MetaLR, MetaRNN, BayesDel, ClinPred, FATHMM, LRT, PROVEAN, DEOGEN2, MVP, VEST4, MPC, PrimateAI, MutationTaster and MutationAssessor \\
AI-derived predictors & AlphaMissense and ESM-derived missense effect scores \\
Engineered biological features & BLOSUM62 scores, amino acid physicochemical changes, codon composition, CpG context, transition/transversion status, consensus voting and interaction features \\
\bottomrule
\end{tabularx}
\end{table}

\newpage
\section*{Supplementary Section 2. ClinVar-labelled benchmark dataset construction}
\addcontentsline{toc}{section}{Supplementary Section 2. ClinVar-labelled benchmark dataset construction}

ClinVar-annotated missense variants were extracted from the chromosome-level merged annotation tables. 
Variants with non-null and non-empty \texttt{CLNSIG} annotations were retained, and clinical significance labels were harmonised into a binary machine-learning target stored as \texttt{ML\_Class}. 
Pathogenic or likely pathogenic variants were assigned to the positive class, whereas benign or likely benign variants were assigned to the negative class. 
Uncertain, conflicting, ambiguous, or unsupported clinical significance categories were excluded from the primary benchmark.

Variant identifiers, including \texttt{Chr}, \texttt{Start}, \texttt{End}, \texttt{Ref}, and \texttt{Alt}, were retained separately for traceability. 
ClinVar-specific fields not intended for prediction, including \texttt{CLNDN}, \texttt{CLNREVSTAT}, \texttt{CLNDISDB}, and \texttt{CLNALLELEID}, were excluded to reduce direct information leakage. 
The resulting ClinVar-labelled dataset was partitioned into five stratified folds, and fold-specific preprocessing was fitted on each training partition before being applied to the corresponding held-out test fold.

\begin{table}[H]
\centering
\caption{ClinVar-labelled benchmark dataset summary.}
\label{tab:s2_clinvar_summary}
\small
\begin{tabular}{lr}
\toprule
\textbf{Item} & \textbf{Value} \\
\midrule
Total ClinVar-labelled missense variants & 132,714 \\
Pathogenic / likely pathogenic variants & 76,804 \\
Benign / likely benign variants & 55,910 \\
Primary cross-validation design & Stratified five-fold cross-validation \\
Primary label encoding & Benign = 0; Pathogenic = 1 \\
Main ranking metric & Matthews correlation coefficient \\
Secondary metric & ROC-AUC \\
\bottomrule
\end{tabular}
\end{table}

\newpage
\section*{Supplementary Section 3. Feature configurations and dataset numbering}
\addcontentsline{toc}{section}{Supplementary Section 3. Feature configurations and dataset numbering}

This section defines the feature configurations used in the primary benchmark and circularity-controlled ablation analyses. 
The numbering below should be used consistently in the main manuscript, tables, figures, and supplementary material.

\begin{table}[H]
\centering
\caption{Dataset numbering and feature configuration definitions.}
\label{tab:s3_dataset_numbering}
\small
\begin{tabularx}{\textwidth}{p{2.2cm}p{3.8cm}X}
\toprule
\textbf{Dataset} & \textbf{Configuration} & \textbf{Definition} \\
\midrule
Dataset 2 & Full benchmark & Full 303-feature benchmark matrix including computational pathogenicity predictors, conservation scores, functional annotations, population-frequency variables, categorical variables, raw biological sequence variables, unified prediction mapping, amino acid property features, codon features, nucleotide substitution features, consensus-voting features, interaction features, AlphaMissense and ESM-derived features. \\
Dataset 3 & Naive feature set & Reduced 41-feature matrix retaining conservation scores, functional annotations, and categorical variables, without the broad pathogenicity predictor and engineered feature space. \\
Dataset 4 & Location-oriented feature set & Reduced 56-feature matrix retaining functional annotations, categorical variables, raw biological sequence variables, and derived biological features such as BLOSUM62 scores, amino acid property changes, codon composition features, and nucleotide substitution features. \\
Dataset 5 & NoPriorPredictors & Circularity-controlled configuration excluding computational pathogenicity predictor scores and population frequency metrics while retaining conservation scores, functional/transcript annotations, categorical variables, and engineered biological features. \\
Dataset 6 & NoClinicalFeatures & Conservative circularity-control configuration excluding features derived from tools trained on ClinVar, HGMD, or overlapping clinical databases, including prior pathogenicity predictors, BayesDel allele-frequency scores, AlphaMissense, ESM-derived scores, MetaRNN, and ClinPred. \\
Dataset 7 & NoAlphaMissenseESM & Configuration retaining the broader prior-predictor feature space but excluding AlphaMissense and ESM-derived features. \\
Dataset 8 & EngineeredBiologicalOnly & Configuration retaining only engineered biological features derived from raw sequence information, including BLOSUM62, amino acid physicochemical changes, codon composition, CpG context, codon bias, transition/transversion status, and nucleotide context. \\
\bottomrule
\end{tabularx}
\end{table}

\subsection*{Section 3.1. Feature filtering and processing}

Feature engineering was implemented as row-independent, fold-specific processing. 
Automatic filtering rules removed features with excessive missingness, constant features, genomic coordinate fields, and ClinVar-derived annotations that could introduce target leakage. 
Categorical variables with manageable cardinality were retained for encoding, whereas unsuitable high-cardinality fields were excluded from standard one-hot encoding or handled separately.

\begin{table}[H]
\centering
\caption{Summary of feature analysis for genome-wide prediction.}
\label{tab:s3_feature_analysis}
\small
\begin{tabular}{lr}
\toprule
\textbf{Feature-analysis item} & \textbf{Value} \\
\midrule
Candidate features evaluated & 334 \\
Chromosome-level files evaluated & 25 \\
Features retained for modelling & 207 \\
Features excluded & 127 \\
Retained numeric features & 151 \\
Retained categorical features & 56 \\
Average missingness among retained features & 14.6\% \\
Median missingness among retained features & 16.4\% \\
\bottomrule
\end{tabular}
\end{table}

\newpage
\section*{Supplementary Section 4. Model training and evaluation}
\addcontentsline{toc}{section}{Supplementary Section 4. Model training and evaluation}

Supervised benchmarking was performed using XGBoost, Random Forest, FLAML AutoML, TabNet, a PyTorch deep neural network, and a TensorFlow deep neural network. 
After fold-specific feature engineering, models were trained independently within each outer cross-validation fold using the processed training partition and evaluated on the corresponding held-out test partition. 
For all machine learning and deep learning pipelines, labels were converted to a binary encoding in which 0 represented benign variants and 1 represented pathogenic variants.

Within each fold, only shared training and test features were retained, missing values were filled with zero, and the inputs were standardised using a \texttt{StandardScaler} fitted on the training partition and applied to the test partition.

\begin{table}[H]
\centering
\caption{Summary of model families evaluated in AnnotateMissense.}
\label{tab:s4_model_families}
\small
\begin{tabularx}{\textwidth}{p{3.2cm}X}
\toprule
\textbf{Model} & \textbf{Training configuration summary} \\
\midrule
XGBoost & 200 trees, maximum depth 8 for the primary benchmark, learning rate 0.1, subsample 0.8, column subsampling 0.8, and fold-specific class weighting using \texttt{scale\_pos\_weight}. \\
Random Forest & 200 trees, maximum depth 15, minimum split size 5, minimum leaf size 2, \texttt{max\_features=sqrt}, bootstrap aggregation, out-of-bag scoring, and balanced class weights. \\
FLAML AutoML & Binary classification with ROC-AUC as the optimisation metric, 600-second time budget per fold, and estimator search over XGBoost, Random Forest, LightGBM, and Extra Trees. \\
TabNet & \texttt{n\_d=64}, \texttt{n\_a=64}, \texttt{n\_steps=7}, \texttt{gamma=1.3}, \texttt{lambda\_sparse=1e-3}, Adam optimiser, entmax masking, and step-wise learning-rate scheduling. \\
PyTorch DNN & Four hidden layers with widths 512, 256, 128, and 64; batch normalisation; dropout rates 0.4, 0.3, 0.2, and 0.1; Adam optimiser; early stopping. \\
TensorFlow DNN & Dense 512--256--128--64 architecture with batch normalisation and dropout, Adam optimisation, binary cross-entropy loss, and internal validation-based early stopping. \\
\bottomrule
\end{tabularx}
\end{table}


\newpage
\section*{Supplementary Section 5. Extended benchmark and ablation results}
\addcontentsline{toc}{section}{Supplementary Section 5. Extended benchmark and ablation results}

Seven supervised feature configurations were evaluated to assess the contribution of different evidence layers to missense pathogenicity prediction. Dataset~2 corresponds to the full 303-feature ClinVar benchmark matrix, Dataset~3 corresponds to the naive 41-feature set, Dataset~4 corresponds to the location-oriented 56-feature set, and Datasets~5--8 correspond to circularity-controlled ablation configurations.

Across all configurations, Dataset~2 produced the strongest performance. XGBoost achieved the highest overall performance on Dataset~2, with a mean test ROC-AUC of 0.99497 and a mean test MCC of 0.94108. FLAML AutoML and Random Forest showed comparable performance, while TabNet and the PyTorch deep neural network also achieved strong performance on the full feature set.

Performance dropped substantially in Dataset~3 and Dataset~4, indicating that restricted feature spaces based mainly on conservation, functional annotation, categorical variables, or local sequence-derived features were insufficient to recapitulate the performance of the full multi-source feature set. The circularity-controlled ablation configurations showed that removing prior predictors and population-frequency features produced a large reduction in MCC, whereas removing AlphaMissense and ESM-derived features alone produced only a minimal reduction.

\begin{landscape}
\begin{table}[H]
\centering
\caption{Unified benchmark performance across Dataset 2--8. Performance is reported as the mean $\pm$ standard deviation across five cross-validation folds, with 95\% confidence intervals in brackets. Dataset numbering follows the manuscript convention: Dataset~2 = full benchmark, Dataset~3 = naive feature set, Dataset~4 = location-oriented feature set, and Datasets~5--8 = circularity-controlled ablation configurations. TensorFlow DNN results are omitted for Dataset~5--8 because of a file caching issue identified post hoc.}
\label{tab:s5_unified_benchmark_corrected}
\scriptsize
\setlength{\tabcolsep}{3pt}
\renewcommand{\arraystretch}{1.05}

\begin{adjustbox}{max width=\linewidth}
\begin{tabular}{clllrll}
\toprule
\textbf{Dataset} & \textbf{Configuration} & \textbf{Type} & 
\textbf{Model} & \textbf{Feats} & 
\textbf{Test ROC-AUC mean $\pm$ sd} & 
\textbf{Test MCC mean $\pm$ sd [95\% CI]} \\
\midrule

\multirow{6}{*}{2} & \multirow{6}{*}{Full benchmark}
  & ML & \textbf{XGBoost}    & 303 & $0.99497 \pm 0.00039$ & $\mathbf{0.94108 \pm 0.00198}$ [0.93862, 0.94354] \\
  & & ML & FLAML AutoML       & 194 & $0.99486 \pm 0.00044$ & $0.94024 \pm 0.00180$ [0.93801, 0.94247] \\
  & & ML & Random Forest      & 194 & $0.99413 \pm 0.00045$ & $0.93638 \pm 0.00153$ [0.93448, 0.93828] \\
  & & DL & TabNet             & 303 & $0.99407 \pm 0.00056$ & $0.93653 \pm 0.00212$ [0.93390, 0.93915] \\
  & & DL & PyTorch DNN        & 303 & $0.99265 \pm 0.00051$ & $0.93508 \pm 0.00338$ [0.93088, 0.93928] \\
  & & DL & TensorFlow DNN     & 303 & $0.99338 \pm 0.00092$ & $0.93118 \pm 0.00303$ [0.92742, 0.93494] \\

\midrule
\multirow{3}{*}{3} & \multirow{3}{*}{Naive feature set}
  & ML & FLAML AutoML        & 41 & $0.82752 \pm 0.00382$ & $\mathbf{0.49889 \pm 0.00763}$ [0.48942, 0.50836] \\
  & & ML & Random Forest      & 41 & $0.82602 \pm 0.00349$ & $0.49675 \pm 0.00734$ [0.48764, 0.50586] \\
  & & ML & XGBoost            & 41 & $0.82398 \pm 0.00366$ & $0.49515 \pm 0.00756$ [0.48577, 0.50453] \\

\midrule
\multirow{6}{*}{4} & \multirow{6}{*}{Location-oriented feature set}
  & ML & Random Forest       & 56 & $0.83115 \pm 0.00273$ & $\mathbf{0.51125 \pm 0.00743}$ [0.50202, 0.52048] \\
  & & ML & XGBoost            & 56 & $0.83082 \pm 0.00199$ & $0.50964 \pm 0.00518$ [0.50320, 0.51608] \\
  & & ML & FLAML AutoML       & 56 & $0.83210 \pm 0.00470$ & $0.49733 \pm 0.00882$ [0.48638, 0.50828] \\
  & & DL & PyTorch DNN        & 56 & $0.83298 \pm 0.00282$ & $0.50910 \pm 0.00574$ [0.50197, 0.51623] \\
  & & DL & TensorFlow DNN     & 56 & $0.83150 \pm 0.00253$ & $0.50854 \pm 0.00443$ [0.50304, 0.51404] \\
  & & DL & TabNet             & 56 & $0.83095 \pm 0.00317$ & $0.49091 \pm 0.00930$ [0.47937, 0.50245] \\

\midrule
\multirow{5}{*}{5} & \multirow{5}{*}{NoPriorPredictors}
  & ML & FLAML AutoML        & 100 & $0.94223 \pm 0.00180$ & $0.73145 \pm 0.00552$ [0.72459, 0.73831] \\
  & & ML & XGBoost            & 100 & $0.93979 \pm 0.00199$ & $0.71977 \pm 0.00598$ [0.71234, 0.72719] \\
  & & ML & Random Forest      & 100 & $0.93698 \pm 0.00218$ & $0.71593 \pm 0.00571$ [0.70884, 0.72302] \\
  & & DL & TabNet             & 100 & $0.94688 \pm 0.02676$ & $\mathbf{0.75824 \pm 0.09961}$ [0.63457, 0.88191] \\
  & & DL & PyTorch DNN        & 100 & $0.94626 \pm 0.02596$ & $0.75518 \pm 0.10097$ [0.62983, 0.88053] \\

\midrule
\multirow{5}{*}{6} & \multirow{5}{*}{NoClinicalFeatures}
  & ML & FLAML AutoML        & 100 & $0.94303 \pm 0.00191$ & $\mathbf{0.73255 \pm 0.00514}$ [0.72616, 0.73893] \\
  & & ML & XGBoost            & 100 & $0.93977 \pm 0.00206$ & $0.71906 \pm 0.00476$ [0.71315, 0.72498] \\
  & & ML & Random Forest      & 100 & $0.93695 \pm 0.00216$ & $0.71640 \pm 0.00630$ [0.70858, 0.72423] \\
  & & DL & TabNet             & 100 & $0.93442 \pm 0.00276$ & $0.71522 \pm 0.00776$ [0.70559, 0.72485] \\
  & & DL & PyTorch DNN        & 100 & $0.93514 \pm 0.00232$ & $0.70993 \pm 0.00629$ [0.70212, 0.71773] \\

\midrule
\multirow{5}{*}{7} & \multirow{5}{*}{NoAlphaMissenseESM}
  & ML & FLAML AutoML        & 302 & $\mathbf{0.99488 \pm 0.00049}$ & $\mathbf{0.94092 \pm 0.00109}$ [0.93957, 0.94227] \\
  & & ML & XGBoost            & 302 & $0.99470 \pm 0.00044$ & $0.94029 \pm 0.00172$ [0.93815, 0.94243] \\
  & & ML & Random Forest      & 302 & $0.99392 \pm 0.00053$ & $0.93489 \pm 0.00149$ [0.93305, 0.93674] \\
  & & DL & TabNet             & 302 & $0.99413 \pm 0.00054$ & $0.93704 \pm 0.00154$ [0.93512, 0.93895] \\
  & & DL & PyTorch DNN        & 302 & $0.99255 \pm 0.00059$ & $0.93495 \pm 0.00208$ [0.93237, 0.93754] \\

\midrule
\multirow{5}{*}{8} & \multirow{5}{*}{EngineeredBiologicalOnly}
  & ML & Random Forest       & 41 & $0.83107 \pm 0.00267$ & $\mathbf{0.51204 \pm 0.00650}$ [0.50397, 0.52010] \\
  & & ML & XGBoost            & 41 & $0.82999 \pm 0.00144$ & $0.50699 \pm 0.00313$ [0.50310, 0.51088] \\
  & & ML & FLAML AutoML       & 41 & $0.83395 \pm 0.00461$ & $0.50057 \pm 0.01079$ [0.48717, 0.51396] \\
  & & DL & PyTorch DNN        & 41 & $0.83259 \pm 0.00266$ & $0.50868 \pm 0.00485$ [0.50267, 0.51470] \\
  & & DL & TabNet             & 41 & $0.83097 \pm 0.00320$ & $0.49185 \pm 0.00887$ [0.48084, 0.50286] \\

\bottomrule
\multicolumn{7}{l}{\scriptsize ML = machine learning; DL = deep learning; MCC = Matthews correlation coefficient.} \\
\multicolumn{7}{l}{\scriptsize Bold values indicate the best MCC within each dataset configuration.} \\
\end{tabular}
\end{adjustbox}
\end{table}
\end{landscape}
\newpage
\section*{Supplementary Section 6. Circularity-controlled ablation results}
\addcontentsline{toc}{section}{Supplementary Section 6. Circularity-controlled ablation results}

Circularity-controlled ablation experiments were performed to assess the contribution of individual feature categories while controlling for potential circularity arising from predictors trained on overlapping clinical variant databases. 
Dataset~5 excluded prior pathogenicity predictors and population frequency metrics. 
Dataset~6 excluded the broader set of features derived from tools with known clinical database overlap. 
Dataset~7 excluded AlphaMissense and ESM-derived scores only. 
Dataset~8 retained only engineered biological sequence-derived features.

\begin{table}[H]
\centering
\caption{Summary of circularity-controlled ablation configurations.}
\label{tab:s6_ablation_definitions}
\small
\begin{tabularx}{\textwidth}{p{2.2cm}p{4.2cm}X}
\toprule
\textbf{Dataset} & \textbf{Configuration} & \textbf{Purpose} \\
\midrule
Dataset 5 & NoPriorPredictors & Tests performance after removing prior pathogenicity predictors and population frequency metrics. \\
Dataset 6 & NoClinicalFeatures & Tests conservative performance after removing tools or scores with potential clinical database overlap. \\
Dataset 7 & NoAlphaMissenseESM & Tests the marginal contribution of AlphaMissense and ESM-derived features over broader annotation features. \\
Dataset 8 & EngineeredBiologicalOnly & Tests signal contained only in sequence-derived engineered biological features. \\
\bottomrule
\end{tabularx}
\end{table}

The ablation results showed that removing all prior pathogenicity predictors and population frequency scores reduced XGBoost performance substantially relative to Dataset~2. 
Removing the broader set of clinically trained tools produced a nearly identical reduction, indicating that the marginal contribution of tools with known ClinVar overlap over and above other annotation scores was limited in this setting. 
In contrast, removing AlphaMissense and ESM-derived scores alone produced a negligible reduction, suggesting that these protein-language-model-derived features did not independently explain performance beyond the broader annotation feature set. 
Restricting the model to engineered biological features only produced performance similar to Dataset~3 and Dataset~4.

\begin{table}[H]
\centering
\caption{Summary interpretation of circularity-controlled ablation findings.}
\label{tab:s6_ablation_interpretation}
\small
\begin{tabularx}{\textwidth}{p{2.2cm}p{3.8cm}X}
\toprule
\textbf{Dataset} & \textbf{Expected effect} & \textbf{Interpretation} \\
\midrule
Dataset 5 & Large performance reduction & Prior pathogenicity predictors and population frequency features contribute strongly to benchmark performance. \\
Dataset 6 & Similar reduction to Dataset 5 & Removing clinically trained tools does not produce a much larger drop than removing prior predictors and frequency features more generally. \\
Dataset 7 & Minimal performance reduction & AlphaMissense and ESM-derived features add limited independent signal after other annotation sources are included. \\
Dataset 8 & Reduced performance similar to Dataset 3/4 & Engineered sequence-derived biological features alone provide useful but insufficient discriminatory signal. \\
\bottomrule
\end{tabularx}
\end{table}

\newpage
\section*{Supplementary Section 7. Feature-importance patterns}
\addcontentsline{toc}{section}{Supplementary Section 7. Feature-importance patterns}

Feature-importance analysis showed that dominant predictive signals depended on the feature configuration. 
In Dataset~2, the highest-ranked variables were integrated pathogenicity confidence scores and established missense effect predictors, complemented by biologically engineered features such as BLOSUM62 score, codon context, and amino acid property changes. 
In Dataset~3, important variables were dominated by conservation and functional consequence annotations. 
In Dataset~4, feature importance shifted toward amino acid chemistry, codon composition, and functional consequence variables.

\begin{table}[H]
\centering
\caption{Representative recurrent important features across interpretable tree-based models.}
\label{tab:s7_recurrent_features}
\small
\begin{tabularx}{\textwidth}{p{4.2cm}p{3.4cm}X}
\toprule
\textbf{Feature} & \textbf{Feature class} & \textbf{Recurrently flagged in} \\
\midrule
\texttt{blosum62\_score} & Amino acid substitution severity & Dataset 2 XGBoost; Dataset 4 XGBoost; Dataset 4 Random Forest \\
\texttt{refcodon\_merged} & Codon context & Dataset 2 XGBoost; Dataset 4 XGBoost; Dataset 4 Random Forest \\
\texttt{ExonicFunc.ccdsGene} & Functional consequence & Dataset 3 Random Forest; Dataset 4 XGBoost; Dataset 4 Random Forest \\
\texttt{Func.knownGene} / \texttt{ExonicFunc.knownGene} & Transcript consequence & Dataset 3 XGBoost; Dataset 3 Random Forest \\
\texttt{phyloP100way\_vertebrate\_dbnsfp42c} & Conservation & Dataset 3 XGBoost; Dataset 3 Random Forest \\
\texttt{GERP++\_RS\_rankscore} & Conservation & Dataset 3 XGBoost; Dataset 3 Random Forest \\
\texttt{blosum\_non\_conservative} & Amino acid severity class & Dataset 4 XGBoost; Dataset 4 Random Forest \\
\texttt{alt\_charged} & Amino acid physicochemical property & Dataset 4 XGBoost; Dataset 4 Random Forest \\
\texttt{aa\_structural\_impact} & Protein-context engineered feature & Dataset 4 XGBoost; Dataset 4 Random Forest \\
\texttt{codon\_gc\_content} & Codon composition & Dataset 4 Random Forest; retained in location-oriented engineered feature set \\
\bottomrule
\end{tabularx}
\end{table}

\newpage
\section*{Supplementary Section 8. Comparison with established pathogenicity predictors}
\addcontentsline{toc}{section}{Supplementary Section 8. Comparison with established pathogenicity predictors}

To compare AnnotateMissense-trained models against established pathogenicity predictors, agreement with ClinVar classifications was evaluated across 135,512 missense variants. 
ClinVar labels were binarised such that pathogenic and likely pathogenic variants were assigned a value of 1, whereas benign and likely benign variants were assigned a value of 0. 
Performance was assessed using Pearson correlation coefficient, Spearman rank correlation coefficient, MCC, and overall classification accuracy. 
For models trained in this study, continuous probability scores were normalised to the range $[0,1]$ before correlation analysis, and binary predictions were derived using a decision threshold of 0.5.

\begin{landscape}
\begin{table}[H]
\centering
\caption{Performance of selected methods relative to ClinVar binary classifications.}
\label{tab:s8_clinvar_comparison}
\small
\begin{tabular}{llccccc}
\toprule
\textbf{Method} & \textbf{Type} & \textbf{Pearson} & \textbf{Spearman} & \textbf{MCC} & \textbf{Accuracy} & \textbf{N} \\
\midrule
\textbf{Dataset 2 XGBoost} & Continuous & 0.983 & 0.853 & 0.981 & 0.991 & 135,512 \\
\textbf{Dataset 2 FLAML AutoML} & Continuous & 0.974 & 0.852 & 0.966 & 0.983 & 135,512 \\
\textbf{Dataset 2 Random Forest} & Continuous & 0.973 & 0.852 & 0.968 & 0.984 & 135,512 \\
MetaRNN & Binary & 0.883 & 0.883 & 0.883 & 0.942 & 101,375 \\
BayesDel\_addAF & Continuous & 0.873 & 0.827 & 0.818 & 0.908 & 133,147 \\
\textbf{Dataset 3 FLAML AutoML} & Continuous & 0.716 & 0.698 & 0.662 & 0.836 & 135,512 \\
\textbf{Dataset 3 Random Forest} & Continuous & 0.704 & 0.690 & 0.645 & 0.827 & 135,512 \\
PROVEAN & Binary & 0.615 & 0.615 & 0.615 & 0.806 & 97,628 \\
PolyPhen-2 HVAR & Binary & 0.582 & 0.582 & 0.582 & 0.785 & 90,798 \\
\textbf{Dataset 3 XGBoost} & Continuous & 0.652 & 0.641 & 0.580 & 0.794 & 135,512 \\
SIFT4G & Binary & 0.556 & 0.556 & 0.556 & 0.774 & 96,662 \\
\textbf{Dataset 4 Random Forest} & Continuous & 0.617 & 0.614 & 0.553 & 0.765 & 135,512 \\
\textbf{Dataset 4 XGBoost} & Continuous & 0.607 & 0.603 & 0.543 & 0.760 & 135,512 \\
SIFT & Binary & 0.538 & 0.538 & 0.538 & 0.753 & 96,573 \\
PolyPhen-2 HDIV & Binary & 0.525 & 0.525 & 0.525 & 0.743 & 90,798 \\
\textbf{Dataset 4 FLAML AutoML} & Continuous & 0.590 & 0.583 & 0.510 & 0.755 & 135,512 \\
MutationTaster & Binary & 0.505 & 0.505 & 0.505 & 0.748 & 133,024 \\
LRT & Binary & 0.455 & 0.455 & 0.455 & 0.742 & 111,973 \\
FATHMM & Binary & 0.441 & 0.441 & 0.441 & 0.722 & 96,726 \\
REVEL & Continuous & 0.793 & 0.766 & 0.426 & 0.670 & 99,337 \\
AlphaMissense & Continuous & 0.792 & 0.764 & 0.420 & 0.658 & 94,209 \\
CADD & Continuous & 0.684 & 0.742 & 0.205 & 0.476 & 133,975 \\
\bottomrule
\end{tabular}
\end{table}
\end{landscape}

\newpage
\section*{Supplementary Section 9. Genome-wide prediction model}
\addcontentsline{toc}{section}{Supplementary Section 9. Genome-wide prediction model}

After feature engineering, per-chromosome feature files were merged across all chromosomes to construct the genome-wide modelling dataset. 
Only variants with a \texttt{CLNSIG} classification of benign or pathogenic were retained for model training and evaluation; variants of uncertain significance or unknown labels were excluded. 
XGBoost was selected as the final classifier for training on the ClinVar-labelled genome-wide dataset.

During preprocessing, multiple representations of missing values were standardised, non-numeric columns were converted to numeric where possible, remaining missing values were replaced with zero, rows containing infinite values were removed, and zero-variance features were discarded. 
The merged dataset was divided into training and test partitions using a stratified 80:20 split. 
SMOTE was applied only to the training data to address class imbalance. 
The saved XGBoost model was then used for genome-wide inference across all engineered per-chromosome variant files.

\begin{table}[H]
\centering
\caption{Final genome-wide XGBoost model configuration.}
\label{tab:s9_final_xgboost}
\small
\begin{tabular}{ll}
\toprule
\textbf{Parameter} & \textbf{Value} \\
\midrule
Model & XGBoost classifier \\
Train/test split & Stratified 80:20 split \\
Class balancing & SMOTE on training data only \\
\texttt{n\_estimators} & 200 \\
\texttt{max\_depth} & 6 \\
\texttt{learning\_rate} & 0.1 \\
\texttt{subsample} & 0.9 \\
\texttt{colsample\_bytree} & 0.9 \\
\texttt{reg\_alpha} & 0.1 \\
\texttt{reg\_lambda} & 1.0 \\
\texttt{random\_state} & 42 \\
Evaluation metric & Log loss \\
Genome-wide inference target & 90,643,830 missense variants \\
\bottomrule
\end{tabular}
\end{table}

\newpage
\section*{Supplementary Section 10. Temporal ClinVar validation}
\addcontentsline{toc}{section}{Supplementary Section 10. Temporal ClinVar validation}

Temporal ClinVar validation was performed to assess generalisation beyond the original ClinVar-derived model-development setting. 
A newer ClinVar \texttt{variant\_summary.txt} file was parsed and restricted to GRCh38 single-nucleotide variants. 
Clinical significance labels were converted to a binary evaluation task using only strict pathogenic and benign categories. 
Variants annotated as uncertain significance, conflicting, ambiguous, or other non-binary categories were excluded from the evaluation.

For each chromosome, variants in the newer ClinVar file were matched to the older annotation-derived ClinVar set by chromosome, VCF position, reference allele, and alternate allele. 
Variants already present in the older set were excluded, leaving newly observed ClinVar pathogenic/benign variants. 
AnnotateMissense predictions were joined to these variants using the same genomic key. 
AnnotateMissense was evaluated using its final categorical prediction, where damaging/pathogenic predictions were encoded as 1 and tolerated/benign predictions were encoded as 0.

\begin{table}[H]
\centering
\caption{Temporal ClinVar validation summary for AnnotateMissense.}
\label{tab:s10_temporal_validation}
\small
\begin{tabular}{lr}
\toprule
\textbf{Metric} & \textbf{Value} \\
\midrule
Newer ClinVar GRCh38 SNVs parsed & 4,129,645 \\
Newly observed strict pathogenic/benign variants & 1,459,305 \\
Newly observed variants with AnnotateMissense categorical predictions & 298,353 \\
MCC & 0.7613 \\
Accuracy & 0.8798 \\
Sensitivity & 0.9070 \\
Specificity & 0.8563 \\
Precision & 0.8453 \\
F1-score & 0.8750 \\
\bottomrule
\end{tabular}
\end{table}

These results support temporal concordance with later ClinVar pathogenic/benign classifications. 
However, they should be interpreted as temporal concordance with ClinVar rather than independent clinical validation, because several input evidence sources and comparator tools may themselves be influenced by ClinVar, HGMD, or related clinical variant databases.


ewpage
\newpage
\section*{Supplementary Section 11. VUS prioritisation and gene-constraint analysis}
\addcontentsline{toc}{section}{Supplementary Section 11. VUS prioritisation and gene-constraint analysis}

To evaluate biological utility beyond aggregate benchmarking, the genome-wide XGBoost classifier was applied to ClinVar missense variants annotated as variants of uncertain significance (VUS; \texttt{Uncertain\_significance}). VUS were intersected with AnnotateMissense genome-wide predictions and assigned a pathogenicity score from the trained classifier. To test whether predicted pathogenicity was consistent with independent gene-level constraint, each VUS was assigned to the gnomAD v2.1.1 missense constraint Z-score (\texttt{mis\_z}) of its associated gene.

Genes were stratified into three constraint groups: missense-intolerant genes with \texttt{mis\_z} $> 3$, intermediate genes with $0 \leq$ \texttt{mis\_z} $\leq 3$, and missense-tolerant genes with \texttt{mis\_z} $< 0$. AnnotateMissense score distributions were summarised using the number of scored VUS, mean score, median score, and the proportion of variants exceeding score thresholds of 0.5, 0.7, and 0.9. High-priority VUS were defined as variants with an AnnotateMissense score exceeding 0.9. Differences between constraint groups were assessed using the Mann--Whitney $U$ test.

AnnotateMissense was applied to 49{,}990 ClinVar missense VUS. VUS in missense-intolerant genes (\texttt{mis\_z} $> 3$; $n = 6{,}786$) received higher predicted pathogenicity scores than VUS in missense-tolerant genes (\texttt{mis\_z} $< 0$; $n = 14{,}864$), with mean scores of 0.631 and 0.548, respectively (Mann--Whitney $U$ test, $p = 1.45 \times 10^{-55}$). An intermediate group of genes ($0 \leq$ \texttt{mis\_z} $\leq 3$; $n = 27{,}283$) showed scores between the two extremes (mean = 0.600; Mann--Whitney $U$ test versus intolerant genes, $p = 2.43 \times 10^{-13}$). The proportion of VUS exceeding a score of 0.9 followed the same gradient: 37.4\% in intolerant genes, 34.0\% in intermediate genes, and 31.5\% in tolerant genes. These results indicate that AnnotateMissense assigns higher predicted pathogenicity to VUS in genes under stronger missense selection constraint, consistent with biological expectation, despite gene identity itself not being used as an explicit training feature.

\begin{table}[H]
\centering
\caption{AnnotateMissense score distribution for ClinVar VUS stratified by gnomAD missense constraint Z-score (\texttt{mis\_z}). High-priority VUS are defined as variants with AnnotateMissense score $> 0.9$.}
\label{tab:s11_vus_constraint}
\small
\begin{tabular}{lrrrrrr}
\toprule
\textbf{Constraint group} & \textbf{N} & \textbf{Mean} & \textbf{Median} & \textbf{\%$>$0.5} & \textbf{\%$>$0.7} & \textbf{\%$>$0.9} \\
\midrule
Missense intolerant (\texttt{mis\_z} $> 3$) & 6{,}786 & 0.631 & 0.748 & 63.9 & 53.0 & 37.4 \\
Intermediate ($0 \leq$ \texttt{mis\_z} $\leq 3$) & 27{,}283 & 0.600 & 0.705 & 61.0 & 50.3 & 34.0 \\
Missense tolerant (\texttt{mis\_z} $< 0$) & 14{,}864 & 0.548 & 0.614 & 55.0 & 45.4 & 31.5 \\
\midrule
\multicolumn{7}{l}{\small Mann--Whitney $U$ test: intolerant vs tolerant, $p = 1.45 \times 10^{-55}$; intolerant vs intermediate, $p = 2.43 \times 10^{-13}$.} \\
\bottomrule
\end{tabular}
\end{table}

\newpage
\section*{Supplementary Section 12. Discordance analysis with established predictors}
\addcontentsline{toc}{section}{Supplementary Section 12. Discordance analysis with established predictors}

A pairwise discordance analysis was performed to evaluate whether AnnotateMissense captures signal complementary to established variant-effect predictors. ClinVar-labelled pathogenic and benign missense variants with available AnnotateMissense predictions were used. AnnotateMissense binary predictions were defined using the trained XGBoost classifier output, with score $\geq 0.5$ classified as pathogenic and score $< 0.5$ classified as benign.

Comparator tools were binarised using commonly applied thresholds: AlphaMissense $\geq 0.5$, REVEL $\geq 0.5$, MetaRNN $\geq 0.5$, BayesDel addAF $\geq 0.0$, CADD PHRED $\geq 15$, ClinPred $\geq 0.5$, VEST4 $\geq 0.5$, MVP $\geq 0.5$, SIFT $\leq 0.05$, and ESM1v combined log-likelihood ratio score, where lower scores indicate greater predicted deleteriousness. For each comparator, variants with available predictions from both AnnotateMissense and the comparator were retained. Concordant variants were defined as those receiving the same binary classification from both tools, whereas discordant variants were defined as those receiving opposing classifications. For each discordant subset, the proportion of variants for which AnnotateMissense agreed with the ClinVar ground-truth label was calculated.

We evaluated pairwise discordance between AnnotateMissense and established pathogenicity predictors using 21{,}462 ClinVar-labelled pathogenic and benign missense variants with available AnnotateMissense predictions, comprising 12{,}505 pathogenic and 8{,}957 benign variants. AnnotateMissense showed broad concordance with established tools, with agreement rates ranging from 64.6\% for ESM1v combined scores to 71.9\% for CADD. Discordance rates were moderate and consistent across comparators, ranging from 28.1\% against CADD to 35.4\% against ESM1v, indicating that AnnotateMissense captures overlapping but non-identical signal relative to existing predictors.

On discordant variants, AnnotateMissense showed the strongest complementary resolution against ESM1v, agreeing with ClinVar labels for 67.9\% of variants where the two methods disagreed. AnnotateMissense also showed partial complementary resolution against MVP, SIFT, and CADD, agreeing with ClinVar labels for 44.0\%, 40.4\%, and 38.2\% of discordant variants, respectively. In contrast, AnnotateMissense showed lower discordant-case resolution against clinically trained meta-predictors such as MetaRNN, BayesDel addAF, and ClinPred, agreeing with ClinVar labels for 9.1\%, 11.3\%, and 7.6\% of discordant variants, respectively. This pattern is expected because these meta-predictors are highly optimised on clinical variant labels and therefore perform strongly on ClinVar-derived ground truth.

ESM1v combined log-likelihood ratio scores showed a strong separation between ClinVar pathogenic and benign variants. Pathogenic variants had substantially lower ESM1v scores than benign variants, with mean scores of $-7.41$ and $-1.24$, respectively, consistent with the expected direction of ESM-derived evolutionary constraint signal. This separation was highly significant using a Mann--Whitney $U$ test ($p = 6.65 \times 10^{-285}$). Together, these results suggest that AnnotateMissense is broadly consistent with established pathogenicity predictors while providing complementary classifications for a subset of variants, particularly relative to protein language model-derived ESM1v scores.

\begin{table}[H]
\centering
\caption{Pairwise discordance between AnnotateMissense and established pathogenicity predictors on ClinVar pathogenic and benign missense variants. Concordant and discordant variants are defined by binary classification at the stated threshold for each comparator. AnnotateMissense correct on discord indicates the percentage of discordant variants for which AnnotateMissense agreed with the ClinVar ground-truth label.}
\label{tab:s12_discordance}
\scriptsize
\setlength{\tabcolsep}{4pt}
\renewcommand{\arraystretch}{1.08}

\begin{adjustbox}{max width=\textwidth}
\begin{tabular}{lrrrr}
\toprule
\textbf{Tool} & 
\textbf{N common} & 
\textbf{Concordant (\%)} & 
\textbf{Discordant (\%)} & 
\textbf{AnnotateMissense correct on discord (\%)} \\
\midrule
AlphaMissense  & 14{,}317 & 69.4 & 30.6 & 21.4 \\
REVEL          & 15{,}431 & 69.7 & 30.3 & 18.4 \\
MetaRNN        & 16{,}030 & 69.1 & 30.9 & 9.1 \\
BayesDel addAF & 20{,}723 & 70.0 & 30.0 & 11.3 \\
CADD           & 20{,}937 & 71.9 & 28.1 & 38.2 \\
ClinPred       & 15{,}727 & 69.8 & 30.2 & 7.6 \\
VEST4          & 20{,}397 & 71.6 & 28.4 & 20.7 \\
MVP            & 14{,}647 & 70.8 & 29.2 & 44.0 \\
SIFT           & 14{,}987 & 67.9 & 32.1 & 40.4 \\
ESM1v combined & 3{,}051  & 64.6 & 35.4 & 67.9 \\
\midrule
\multicolumn{5}{p{0.98\textwidth}}{\scriptsize 
Thresholds: AnnotateMissense, AlphaMissense, REVEL, MetaRNN, ClinPred, VEST4, and MVP $\geq 0.5$; BayesDel addAF $\geq 0.0$; CADD PHRED $\geq 15$; SIFT $\leq 0.05$; ESM1v lower scores indicate greater predicted deleteriousness.} \\
\bottomrule
\end{tabular}
\end{adjustbox}
\end{table}


ewpage
\section*{Supplementary Section 13. Data, code, and resource availability}
\addcontentsline{toc}{section}{Supplementary Section 13. Data, code, and resource availability}

The AnnotateMissense workflow, command files, feature-processing scripts, benchmarking scripts, temporal validation scripts, biological use-case analyses, and genome-wide inference scripts are provided through the project repository. The generated genome-wide prediction database and compressed prediction outputs are provided as a public research resource through Zenodo.

\begin{table}[H]
\centering
\caption{Resource availability summary.}
\label{tab:s13_resource_availability}
\small
\begin{tabularx}{\textwidth}{p{4cm}X}
\toprule
\textbf{Resource} & \textbf{Description} \\
\midrule
Main code repository & AnnotateMissense workflow scripts, command files, annotation processing, feature engineering, benchmarking, temporal validation, biological use-case analyses, and genome-wide inference scripts. Available at \url{https://github.com/MuhammadMuneeb007/CAGI7_Annotate_All_Missense}. \\
Genome-wide prediction output & AnnotateMissense pathogenicity scores and binary predictions for 90,643,830 hg38 missense variants. Available from Zenodo at \url{https://doi.org/10.5281/zenodo.19981867}. \\
DuckDB database & Queryable compressed database containing genome-wide AnnotateMissense predictions and selected variant annotations. Available from Zenodo at \url{https://doi.org/10.5281/zenodo.19981867}. \\
Zenodo deposition & Public data deposition for the generated database and prediction output files. DOI: \url{https://doi.org/10.5281/zenodo.19981867}. \\
Supplementary Material 2 & Machine-readable workbook containing database information, feature information, prediction correlation and comparator results, and temporal ClinVar validation results. \\
\bottomrule
\end{tabularx}
\end{table}

\section*{Notes}

This supplementary file provides extended methodological details, benchmark results, ablation analyses, validation procedures, biological use-case analyses, and supporting resource information for the AnnotateMissense manuscript. AnnotateMissense is intended for research prioritisation, benchmarking, and variant triage, and should not be interpreted as a standalone clinical classification system.